\documentclass[conference]{IEEEtran}

\usepackage[pdftex]{graphicx}

\usepackage{mathtools}
\usepackage{algorithm}
\usepackage{algpseudocode}
\DeclareMathOperator{\sgn}{sgn}

\begin{document}

\title{Crafting Adversarial Input Sequences\\for Recurrent Neural Networks}

% author names and affiliations
% use a multiple column layout for up to three different
% affiliations
\author{\IEEEauthorblockN{Nicolas Papernot and Patrick McDaniel}
\IEEEauthorblockA{The Pennsylvania State University\\University Park, PA\\
\{ngp5056,mcdaniel\}@cse.psu.edu}
\and
\IEEEauthorblockN{Ananthram Swami and Richard Harang}
\IEEEauthorblockA{United States Army Research Laboratory\\Adelphi, MD\\
\{ananthram.swami,richard.e.harang\}.civ@mail.mil}
}

% use for special paper notices
%\IEEEspecialpapernotice{(Invited Paper)}

% make the title area
\maketitle

% As a general rule, do not put math, special symbols or citations
% in the abstract
\begin{abstract}
Machine learning models are frequently used to solve complex  security problems, as well as to make decisions in sensitive situations like guiding autonomous vehicles or predicting financial market behaviors. Previous efforts have shown that numerous machine learning models were vulnerable to adversarial manipulations of their inputs taking the form of adversarial samples. Such inputs are crafted by adding carefully selected perturbations to legitimate inputs so as to force the machine learning model to misbehave, for instance by outputting a wrong class if the machine learning task of interest is classification. In fact, to the best of our knowledge, all previous work on adversarial samples crafting for neural network considered models used to solve classification tasks, most frequently in computer vision applications. In this paper, we contribute to the field of adversarial machine learning by investigating adversarial input sequences for recurrent neural networks processing sequential data. We show that the classes of algorithms introduced previously to craft adversarial samples misclassified by feed-forward neural networks can be adapted to recurrent neural networks. In a experiment, we show that adversaries can craft adversarial sequences misleading both  categorical and sequential recurrent neural networks.
\end{abstract}

% no keywords

\IEEEpeerreviewmaketitle

\section{Introduction}
\label{sec:introduction}

Efforts in the machine learning~\cite{szegedy2013intriguing,goodfellow2014explaining} and security~\cite{papernot2015limitations,papernot2016practical} communities have uncovered the vulnerability of machine learning models to adversarial manipulations of their inputs. Specifically, approximations made by training algorithms as well as the underlying linearity of numerous machine learning models, including neural networks, allow adversaries to compromise the integrity of their output using crafted perturbations. Such perturbations are carefully selected to be small---they are often indistinguishable to humans---but at the same time yield important changes of the output of the machine learning model. Solutions making models more robust to adversarial perturbations have been proposed in the literature~\cite{szegedy2013intriguing,goodfellow2014explaining,papernot2015distillation,WardeFarley16}, but models remain largely vulnerable. The existence of this threat vector puts machine learning models at risk when deployed in potentially adversarial settings~\cite{mpc16}. 

A taxonomy of attacks against deep learning classifiers is introduced in~\cite{papernot2015limitations}. To select perturbations changing the class (e.g., label) assigned by a neural network classifier to any class different from the legitimate class~\cite{goodfellow2014explaining} or a specific target class chosen by the adversary~\cite{szegedy2013intriguing,papernot2015limitations}, two approaches can be followed: the \emph{fast gradient sign method}~\cite{goodfellow2014explaining} and the \emph{forward derivative} method~\cite{papernot2015limitations}. Both approaches estimate the model's sensitivity by differentiating  functions defined over its architecture and parameters. The approaches differ in perturbation selection. These techniques were primarily evaluated on models trained to solve image classification tasks. Such tasks simplify adversarial sample crafting because model inputs use linear and differentiable pre-processing: images encoded as numerical vectors. Thus,  perturbations found for the model's input are easily transposed in the corresponding raw image. On the contrary, we study adversarial samples for models mapping sequential inputs pre-processed in a non-linear and non-differentiable manner with categorical or sequential outputs. 

Recurrent Neural Networks (RNNs) are machine learning models adapted from feed-forward neural networks to be suitable for learning mappings between sequential inputs and outputs~\cite{rumelhart1988learning}. They are, for instance, powerful models for sentiment analysis, which can serve the intelligence community in performing analysis of communications in terrorist networks. Furthermore, RNNs can be used for malware classification~\cite{pascanu2015malware}. Predicting sequential data also finds applications in stock analysis for financial market trend prediction. Unlike feed-forward neural networks, RNNs are capable of handling sequential data of large---and often variable---length. RNNs introduce cycles in their computational graph to efficiently model the influence of time~\cite{Goodfellow-et-al-2016-Book}. The presence of cyclical computations potentially presents challenges  to the applicability of existing adversarial sample algorithms based on model differentiation, as cycles prevent computing gradients directly by applying the chain rule. This issue was left as future work by previous work~\cite{papernot2015limitations}.

This is precisely the question we investigate in this paper. We study a particular instance of adversarial examples---which we refer to as \emph{adversarial sequences}---intended to mislead RNNs into producing erroneous outputs. We show that the \emph{forward derivative}~\cite{papernot2015limitations} can be adapted to neural networks with cyclical computational graphs, using a technique named \emph{computational graph unfolding}. In an experiment, we demonstrate how using this forward derivative, i.e. model Jacobian, an adversary can produce adversarial input sequences manipulating both the sequences output by a sequential RNN and classification predictions made by a categorical RNN. Such manipulations do not require the adversary to alter any part of the model's training process or data. In fact, perturbations instantly manipulate the model's output at test time, after it is trained and deployed to make predictions on new inputs.

\newpage

\noindent The contributions of this paper are the following:
\begin{itemize}
\item We formalize the adversarial sample optimization problem in the context of sequential data. We adapt crafting algorithms using the forward derivative to the specificities of RNNs. This includes showing how to compute the forward derivative for cyclical computational graphs. 
\item We investigate transposing adversarial perturbations from the model's pre-processed inputs to the raw inputs.
\item We evaluate the performance of our technique using RNNs making categorical and sequential predictions. On average, changing $9$ words in a $71$ word movie review is sufficient for our categorical RNN to make $100\%$ wrong class predictions when performing sentiment analysis on reviews. We also show that sequences crafted using the Jacobian perturb the sequential outputs of a second RNN.  
\end{itemize}

This paper is intended as a presentation of our initial efforts in an on-going line of research. We include a discussion of future work relevant to the advancement of this research topic.

\section{About Recurrent Neural Networks}
\label{sec:rnn-background}

To facilitate our discussion of adversarial sample crafting techniques in Section~\ref{sec:crafting-adv-sequences}, we provide here an overview of neural networks and more specifically of recurrent neural networks, along with examples of machine learning applications and tasks that can be solved using such models.

\vspace*{0.1in}
\noindent\textbf{Machine Learning -} Machine learning provides automated methods for the analysis of large sets of data~\cite{murphy2012machine}. Tasks solved by machine learning are generally divided in three broad types: \emph{supervised learning}, \emph{unsupervised learning}, and \emph{reinforcement learning}. When the method is designed to learn a mapping (i.e. association) between inputs and outputs, it is an instantiation of \emph{supervised learning}. In such settings, the output data nature characterizes varying problems like classification~\cite{krizhevsky2012imagenet, dahl2013large, cirecsan2012multi}, pattern recognition~\cite{bishop2006pattern}, or regression~\cite{freedman2009statistical}. When the method is only given unlabeled inputs, the machine learning task falls under \emph{unsupervised learning}. Common applications include dimensionality reduction or network pre-training. Finally, \emph{reinforcement learning} considers agents maximizing a reward by taking actions in an environment. Interested readers are referred to the presentation of machine learning in~\cite{murphy2012machine}.

\vspace*{0.05in}
\noindent\textbf{Neural Networks -} Neural Networks are a class of machine learning models that are useful across all tasks of supervised, unsupervised and reinforcement learning. They are made up of neurons---elementary computing units---applying \emph{activation functions} $\phi$ to their inputs $\vec{x}$ in order to produce outputs typically processed by other neurons. The computation performed by a neuron thus takes the following formal form: 
\begin{equation}
\label{eq:ff-neuron}
h(\vec{x}) = \phi(\vec{x}, \vec{w})
\end{equation} 
where $\vec{w}$ is a parameter, referred to as the weight vector, whose role is detailed below. In a neural network $f$, neurons are typically grouped in inter-connected \emph{layers} $f_k$. A network always has at least two layers corresponding to the \emph{input} and \emph{output} of the model. One or more intermediate \emph{hidden} layers can be inserted between these input and output layers. If the network possesses one or no hidden layer, it is referred to as a \emph{shallow neural network}. Otherwise, the network is said to be \emph{deep} and the common interpretation of the hidden layers is that they extract successive and hierarchical representations of the input required to produce the output~\cite{Goodfellow-et-al-2016-Book}. Neural networks are principally parameterized by the weights placed on links between neurons. Such weight parameters $\theta$ hold the model's knowledge and their values are learned during training by considering collections of inputs $\vec{x}$ (with their corresponding labels $y$ in the context of supervised learning).

\begin{figure}[t] 
	\centering
	\includegraphics[width=0.5\columnwidth]{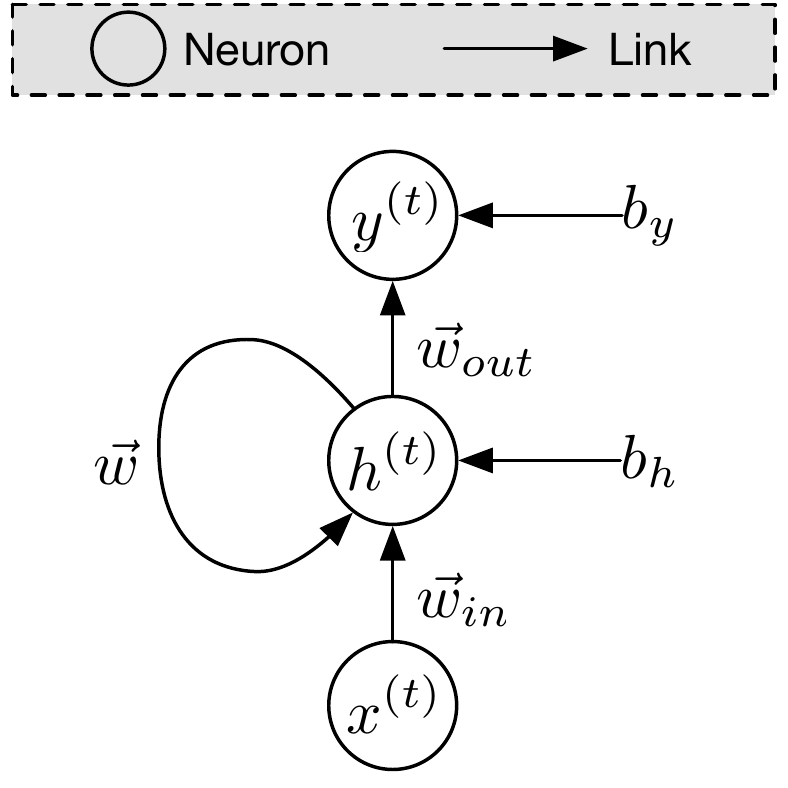} 
	\caption{\textbf{Recurrent Neural Network}: the sequential input $\vec{x}$ is processed by time step value $x^{(t)}$. The hidden neuron evaluates its state $h^{(t)}$ at time step $t$ by adding  (1) the result of multiplying the current input value $x^{(t)}$ with weight $\vec{w}_{in}$, with (2)  the result of multiplying its previous state with weight $\vec{w}$, and (3) the bias $b_h$, and finally applying the hyperbolic tangent. The output $y^{(t)}$ multiplies the hidden neuron state by weight $\vec{w}_{out}$ and adds bias $b_y$.}
	\label{fig:rnn}
\end{figure}

\vspace*{0.05in}
\noindent\textbf{Recurrent Neural Networks -} Recurrent Neural Networks (RNNs) are a variant of the \emph{vanilla} networks described above that is adapted to the modeling of sequential data~\cite{rumelhart1988learning}. Without such sequence-based specificities, vanilla neural networks do not offer the scalability required for the modeling of large sequential data~\cite{Goodfellow-et-al-2016-Book}. The specificities of recurrent neural networks include most importantly the introduction of \emph{cycles} in the model's computational graph, which results in a form of \emph{parameter sharing} responsible for the scalability to large sequences. In other words, in addition to the links between neurons in different layers, recurrent neural networks allow for links between neurons co-located in the same layer, which results in the presence of cycles in the network's architecture. Cycles allow the model to share the weights---which are parameters of the links connecting neuron outputs and inputs---throughout successive values of a given input value at different time steps. In the case of RNNs, Equation (\ref{eq:ff-neuron}) thus becomes: 
\begin{equation}
\label{eq:rnn-neuron}
h^{(t)}(\vec{x}) = \phi\left(h^{(t-1)}(\vec{x}), \vec{x}, \vec{w}\right)
\end{equation} 
following the notation introduced in~\cite{Goodfellow-et-al-2016-Book} where $h^{(t)}(\vec{x})$ is the neuron output---also named state---at time step $t$ of the input sequence. Note that the cycle allows for the activation function to take into account the state of the neuron at the previous time step $t-1$. Thus, the state can be used to transfer some aspects of the previous sequence time steps to upcoming time steps. An example recurrent neural network architecture---used throughout Sections~\ref{sec:crafting-adv-sequences} and~\ref{sec:evaluation}---is illustrated in Figure~\ref{fig:rnn}.

\section{Crafting Adversarial Sequences}
\label{sec:crafting-adv-sequences}

\noindent In the following, we formalize adversarial sequences. We then build on techniques designed to craft adversarial samples for neural network classifiers and adapt them to the problem of crafting adversarial sequences for recurrent neural networks.

\subsection{Adversarial Samples and Sequences}

\vspace*{0.05in}
\noindent\textbf{Adversarial Samples -} In the context of a machine learning classifier $f$, an adversarial samples $\vec{x^*}$ is crafted from a legitimate sample $\vec{x}$ by selecting the smallest---according to a norm appropriate for the input domain---perturbation $\delta_{\vec{x}}$ which results in the altered sample $\vec{x^*}$ being misclassified in a class different from its legitimate class $f(\vec{x})$. The adversarial target class can be a chosen class~\cite{szegedy2013intriguing,papernot2015limitations} or any class different from the legitimate class~\cite{goodfellow2014explaining}. Thus, an adversarial sample solves the following optimization problem, first formalized in~\cite{szegedy2013intriguing}:
\begin{equation}
\label{eq:adv-sample-opt-pb}
\vec{x^*}=\vec{x}+\delta_{\vec{x}}=\vec{x}+\min \| \vec{z}\| \ \mathtt{ s.t. }\ f(\vec{x}+\vec{z}) \neq f(\vec{x}) 
\end{equation}
in the case where the adversary is interested in any target class different from the legitimate class. Finding an exact solution to this problem is not always possible, especially in the case of deep neural networks, due to their non-convexity and non-linearity. Thus, previous efforts introduced methods---two are discussed below---to find approximative solutions~\cite{szegedy2013intriguing,goodfellow2014explaining,papernot2015limitations}. 

\vspace*{0.05in}
\noindent\textbf{Adversarial Sequences -} Consider RNNs processing sequential data. When both the input and output data are sequences, as is the case in one of our experiments, Equation (\ref{eq:adv-sample-opt-pb}) does not hold as the output data is not categorical. Thus, the adversarial sample optimization problems needs to be generalized to specify an adversarial target vector $\vec{y^*}$, which is to be matched as closely as possible by model $f$ when processing the adversarial input $\vec{x^*}$. This can be stated as: 
\begin{equation}
\label{eq:rnn-adv-sample-opt-pb}
\vec{x^*}=\vec{x}+\delta_{\vec{x}}=\vec{x}+\min \| \vec{z}\| \ \mathtt{ s.t. }\   \| f(\vec{x}+\vec{z}) - \vec{y^*} \| < \Delta 
\end{equation}
where $\vec{y^*}$ is the output sequence desired by the adversary, $\| \cdot \|$ a norm appropriate to compare vectors in the RNN's input or output domain, and $\Delta$ the acceptable error between the model output $f(\vec{x}+\vec{z})$ on the adversarial sequence and the adversarial target $\vec{y^*}$. An example norm to compare input sequences is the number of sequence steps perturbed. We detail how approximative solutions---adversarial sequences--- to this problem can be found by computing the model's Jacobian.

\subsection{Using the Fast Gradient Sign Method}

The \emph{fast gradient sign method} approximates the problem in Equation (\ref{eq:adv-sample-opt-pb}) by linearizing the model's cost function around its input and selecting a perturbation using the gradient of the cost function with respect to the input itself~\cite{goodfellow2014explaining}. This gradient can be computed by following the steps typically used for back-propagation during training, but instead of computing gradients with respect to the model parameters (with the intent of reducing the prediction error) as is normally the case during training, the gradients are computed with respect to the input. This yields the following formulation of adversarial samples:
\begin{equation}
\label{eq:fgsm}
\vec{x^*}=\vec{x}+\delta_{\vec{x}} =\vec{x}+ \varepsilon \sgn (\nabla_{\vec{x}} c(f, \vec{x}, \vec{y}))
\end{equation}
where $c$ is the cost function associated with model $f$ and $\varepsilon$ a parameter controlling the perturbation's magnitude. Increasing the input variation parameter $\varepsilon$ increases the likeliness of $\vec{x^*}$ being misclassified but albeit simultaneously increases the perturbation's magnitude and therefore its distinguishability.

As long as the model is differentiable, the fast gradient sign method still applies---even if one inserts recurrent connections in the computational graph of the model. In fact, Goodfellow et al. used the method in~\cite{goodfellow2014explaining} to craft adversarial samples on a multi-prediction deep Boltzmann machine~\cite{goodfellow2013multi}, which uses recurrent connections to classify inputs of fixed size. The adversarial sample crafting method described in Equation (\ref{eq:fgsm}) can thus be used with recurrent neural networks, as long as their loss is differentiable and their inputs 
continuous-valued. We are however also interested in solving Equation (\ref{eq:rnn-adv-sample-opt-pb}) for a model $f$ processing non-continuous input sequence steps.  

\subsection{Using the Forward Derivative}

The forward derivative, introduced in~\cite{papernot2015limitations}, is an alternative means to craft adversarial samples. The method's design considers the threat model of adversaries interested in misclassifying samples in chosen adversarial targets. Nevertheless, the technique can also be used to achieve the weaker goal of misclassification in any target class different from the original sample's class. The forward derivative is defined as the model's Jacobian: 
\begin{equation}
\label{eq:fw-der}
J_f[i,j]= \frac{\partial f_j}{\partial x_i} 
\end{equation}
where $x_i$ is the $i^{th}$ component of the input and $f_j$ the $j^{th}$ component of the output. It precisely evaluates the sensitivity of output component $f_j$ to the input component $x_i$, i.e. it gives a quantified understanding of how input variations modify the output's value by input-output component pair. 

\begin{figure}[t] 
	\centering
	\includegraphics[width=0.7\columnwidth]{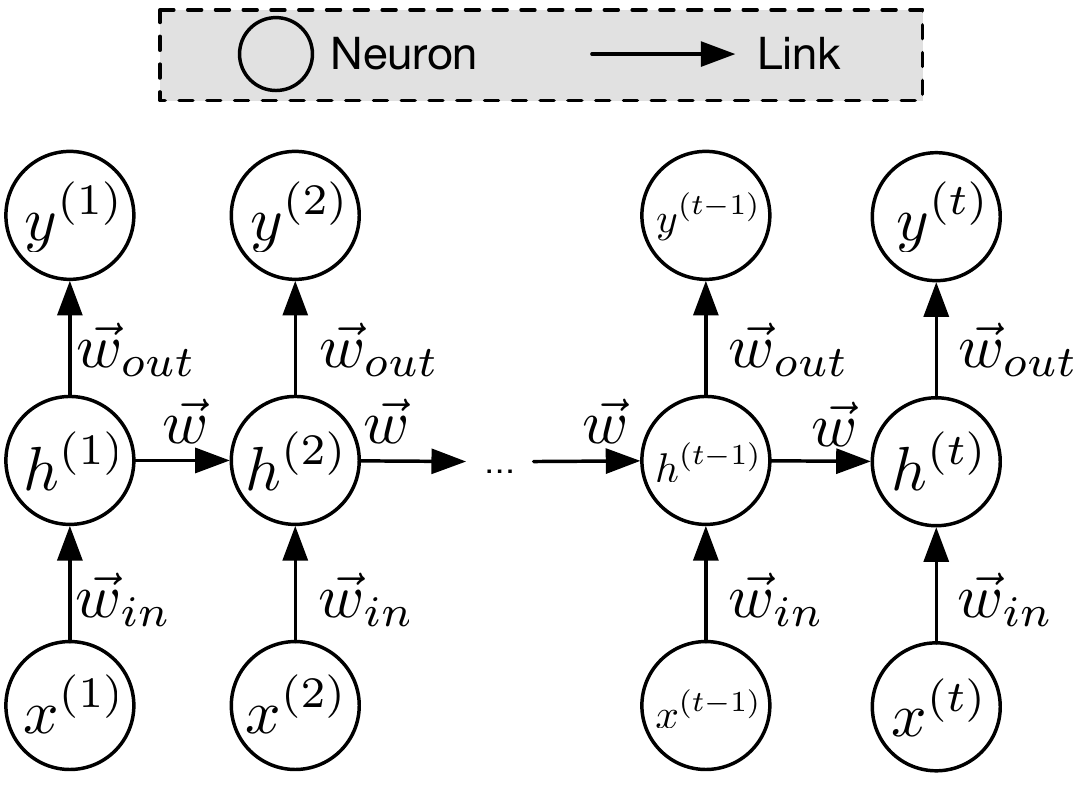} 
	\caption{\textbf{Unfolded Recurrent Neural Network}: this neural network is identical to the one depicted in Figure~\ref{fig:rnn}, with the exception of its recurrence cycle, which is now unfolded. Biases are omitted for clarity of the illustration.}
	\label{fig:rnn-unfolded}
\end{figure}

We leverage the technique known as \emph{computational graph unfolding}~\cite{mozer1989focused,werbos1988generalization} to compute the forward derivative in the presence of cycles, as is the case with RNNs. Looking back at Equation (\ref{eq:rnn-neuron}), one can observe that to compute the neuronal state at time step $t$, we can recursively apply the formula while decrementing the time step. This yields the following:
\begin{equation}
\label{eq:rnn-neuron-unfolded}
h^{(t)}(\vec{x}) = \phi\left(\phi\left(...\ \phi\left(h^{(1)}(\vec{x}), \vec{x}, \vec{w}\right), ...\ \vec{x}, \vec{w}\right), \vec{x}, \vec{w}\right)
\end{equation} 
which is the unfolded version of Equation (\ref{eq:rnn-neuron}). Thus, by unfolding its recurrent components, the computational graph of a recurrent neural network can be made acyclic. For instance, Figure~\ref{fig:rnn-unfolded} draws the unfolded neural network corresponding to the RNN originally depicted in Figure~\ref{fig:rnn}. Using, this unfolded version of the graph, we can compute the recurrent neural network's Jacobian. It can be defined as the following matrix: 
\begin{equation}
\label{eq:fw-der-rnn}
J_f[i,j]= \frac{\partial y^{(j)}}{\partial x^{(i)}}
\end{equation}
where $x^{(i)}$ is the step $i$ of input sequence $\vec{x}$, $y^{(j)}$ is the step $j$ of output sequence $\vec{y}$, and $(i,j)\in [1..t]^2$ for input and output sequences of length $t$. Using the definition of $y^{(j)}$, we have:
\begin{eqnarray}
& \frac{\partial y^{(j)}}{\partial x^{(i)}} = \frac{\partial \phi \left(\vec{w}_{out} \cdot  h^{(j)}+b_y \right)}{\partial x^{(i)}} \nonumber\\
= & \frac{\partial \phi \left(\vec{w}_{out} \cdot  \phi \left(\vec{w} \cdot  h^{(j-1)} + \vec{w}_{in} \cdot  x^{(j)} +b_h \right)+b_y \right)}{\partial x^{(i)}} \nonumber\\
= & \frac{\partial \phi \left(\vec{w}_{out} \cdot  \phi \left(\vec{w} \cdot  \phi \left(\vec{w} \cdot  h^{(j-2)} + \vec{w}_{in} \cdot  x^{(j-1)} +b_h \right) + \vec{w}_{in} \cdot  x^{(j-1)} +b_h \right)+b_y \right)}{\partial x^{(i)}} \nonumber
\end{eqnarray}
By unfolding recursively each time step of the hidden neuron's state until we reach $j-(j-1)=1$, we can write:
\begin{eqnarray}
\label{eq:fw-der-rnn}
& \frac{\partial y^{(j)}}{\partial x^{(i)}} = \frac{\partial \phi \left(\vec{w}_{out} \cdot \phi\left(...\ \phi \left(\vec{w} \cdot  h^{(1)} + \vec{w}_{in} \cdot  x^{(1)} +b_h \right) ...\ \right) +b_y \right)}{\partial x^{(i)}}& 
\end{eqnarray}
which can be evaluated using the chain-rule, as demonstrated by~\cite{papernot2015limitations} in the context of feed-forward neural networks. 

We can craft adversarial sequences for two types of RNN models---\emph{categorical} and \emph{sequential}---with the forward derivative. Previous work introduced adversarial saliency maps to select perturbations using the forward derivative in the context of multi-class classification neural networks~\cite{papernot2015limitations}. Due to space constraints, we do not include an overview of saliency maps because we study a binary classifier in Section~\ref{sec:evaluation}, thus simplifying perturbation selection. Indeed perturbing an input to reduce one class probability necessarily increases the probability given to the second class. Thus, adversarial sequences are crafted by solely considering the Jacobian $J_f[:,j]$ column corresponding to one of the output components $j$.

We now consider crafting adversarial sequences for models outputting sequences. To craft an adversarial sequence $\vec{x^*}$ from a legitimate input sequence $\vec{x}$, we need to select a perturbation $\delta_{\vec{x}}$ such that $f(\vec{x^*})$ is within an acceptable margin of the desired adversarial output $\vec{y^*}$, hence approximatively solving Equation (\ref{eq:rnn-adv-sample-opt-pb}). Consider the output sequence step-by-step: each Jacobian's column corresponds to a step $j$ of the output sequence. We identify a subset of input components $i$ with high absolute values in this column and comparably small absolute values in the other columns of the Jacobian matrix. These components will have a large impact on the RNN's output at step $j$ and a limited impact on its output at other steps. Thus, if we modify components $i$ in the direction indicated by $\sgn(J_f[i,j])\times \sgn(\vec{y^*_j})$, the output sequence's step $j$ will approach the desired adversarial output's component $j$. This method is evaluated in the second part of Section~\ref{sec:evaluation}.

\section{Evaluation}
\label{sec:evaluation}

We craft adversarial sequences for \emph{categorical} and \emph{sequential} RNNs. The categorical RNN performs a sentiment analysis to classify movie reviews (in lieu of intelligence reports) as positive or negative. We mislead this classifier by altering words of the review. The second RNN is trained to learn a mapping between synthetic input and output sequences. The Jacobian-based attack alters the model's output by identifying the contribution of each input sequence step. 

\subsection{Recurrent Neural Networks with Categorical Output}

This RNN is a movie review classifier. It takes as an input a sequence of words---the review---and performs a sentiment analysis to classify it as negative (outputs $0$) or positive (outputs $1$). We were able to achieve an error rate of $100\%$ on the training set by changing on average $9.18$ words in each of the $2,000$ reviews, which are on average $71.06$ word long.

\begin{figure}[h] 
	\centering
	\includegraphics[width=\columnwidth]{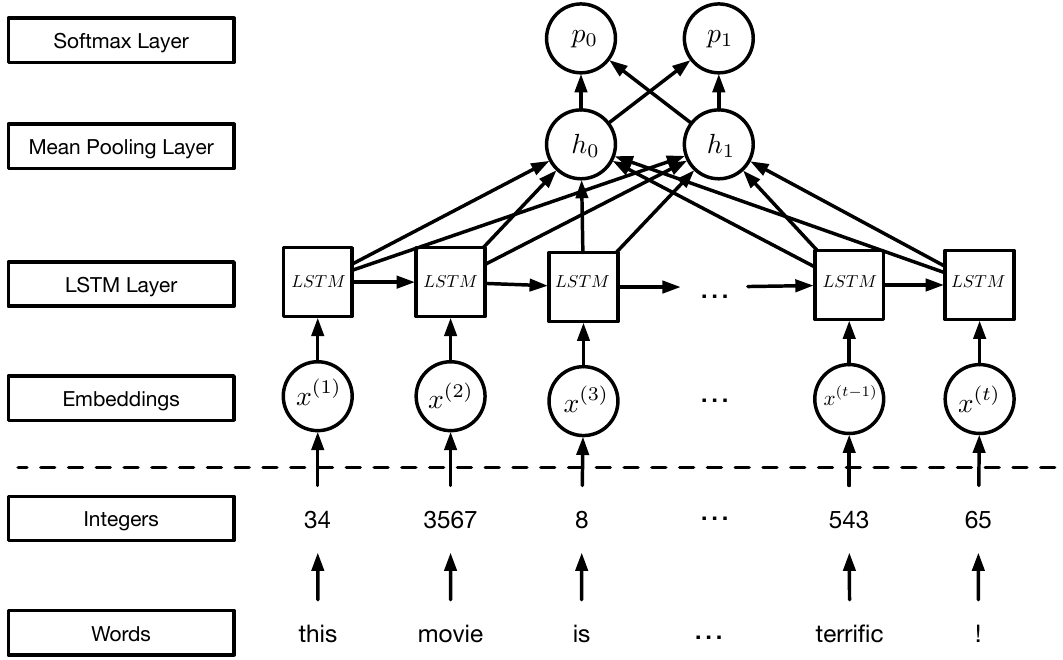} 
	\caption{\textbf{LSTM-based RNN:} this recurrent model classifies movie reviews.}
	\label{fig:lstm}
\end{figure}

\vspace*{0.1in}
\noindent\textbf{Experimental Setup -} We experiment with the \emph{Long Short Term Memory} (LSTM) RNN architecture~\cite{hochreiter1997long}. LSTMs  prevent exploding and vanishing gradients at training by introducing a memory cell, which gives more flexibility to the self-recurrent connections compared to a vanilla RNN, allowing it to remember or forget previous states. Our RNN is composed of four layers---input, LSTM,  mean pooling, and softmax---as shown in Figure~\ref{fig:lstm}. The mean pooling layer averages representations extracted by memory cells of the LSTM layer while the softmax formats the output as probability vectors. 

The RNN $f$ is implemented in Python with Theano~\cite{bergstra2010theano} to facilitate symbolic gradient computations. We train using a little over $2,000$ training and $500$ testing reviews~\cite{maas-EtAl:2011:ACL-HLT2011}. Reviews are sequences of words from a dictionary $D$ that includes $10,000$ words frequently used in the reviews and a special keyword for all other words. The dictionary maps words to integer keys. We convert these integer sequences to matrices, where each row encodes a word as a set of $128$ coordinates---known as \emph{word embeddings}~\cite{hinton1986learning,mesnil2013investigation}. The matrices are used as the input to the RNN described above. Once trained, the architecture achieves accuracies of $100\%$ and $78.21\%$ respectively on the training and testing tests.

\begin{algorithm}[t]
\caption{\textbf{Adversarial Sequence Crafting for the LSTM model:} the algorithm iteratively modifies words $i$ in the input sentence $\vec{x}$ to produce an adversarial sequence $\vec{x^*}$ misclassified by the LSTM architecture $f$ illustrated in Figure~\ref{fig:lstm}.}
\label{alg:adv-seq-crafting}
\begin{algorithmic}[1]
	\Require $f$, $\vec{x}$, $D$
	\State $y \leftarrow f(\vec{x})$
	\State $\vec{x^*} \leftarrow \vec{x}$
	\While{$f(\vec{x^*}) == y$}
		\State Select a word $i$ in the sequence $\vec{x^*}$
		\State $\vec{w} = \| \arg\min_{\vec{z}\in D} \sgn\left(\vec{x^*}[i] - \vec{z}\right) - \sgn(J_f(\vec{x})[i,y])\|$
		\State $\vec{x^*}[i] \leftarrow \vec{w}$
	\EndWhile
	\State \Return $\vec{x^*}$
\end{algorithmic}
\end {algorithm}

\vspace*{-0.3in}

\noindent\textbf{Adversarial Sequences -} We now demonstrate how adversaries can craft adversarial sequences, i.e. sentences misclassified by the model. Thus, we need to identify dictionary words that we can use to modify the sentence $\vec{x}$ in a way that switches its predicted class from positive to negative (or vice-versa). We turn to the attack described in Section~\ref{sec:crafting-adv-sequences} based on computing the model's Jacobian. We evaluate the Jacobian tensor\footnote{The Jacobian is a tensor and not a matrix because each word embedding is a vector itself, so $J_f(\vec{x})[i,j]$ is also a vector and $J_f$ has three dimensions.}  with respect to the embedding inputs: $J_f(\vec{x})[i,j]=\frac{\partial h_j}{\partial x^{(i)}}$. This gives us a precise mapping between changes made to the word embeddings and variations of the output of the pooling layer.\footnote{As indicated in~\cite{papernot2015limitations}, we consider the logits---input values---of the softmax layer instead of its output probabilities to compute the Jacobian because the gradient computations are more stable and the results are the same: the maximum logit index corresponds to the class assigned to the sentence.}  For each word $i$ of the input sequence, $\sgn(J_f(\vec{x})[i,f(\vec{x})])$ where $f(\vec{x})=\arg\max_{0,1}(p_j)$ gives us the direction in which we have to perturb each of the word embedding components in order to reduce the probability assigned to the current class, and thus change the class assigned to the sentence. 

Unlike previous efforts describing adversarial samples in the context of computer vision~\cite{szegedy2013intriguing,goodfellow2014explaining,papernot2015limitations}, we face a difficulty: the set of legitimate word embeddings is finite. Thus, we cannot set the word embedding coordinates to any real value in an adversarial sequence $\vec{x^*}$. To overcome this difficulty, we follow the procedure detailed in Algorithm~\ref{alg:adv-seq-crafting}. We find the word $\vec{z}$ in dictionary $D$ such that the sign of the difference between the embeddings of $\vec{z}$ and the original input word is closest to $\sgn(J_f(\vec{x})[i,f(\vec{x})])$. This embedding takes the direction closest to the one indicated by the Jacobian as most impactful on the model's prediction. By iteratively applying this heuristic to sequence words, we eventually find an adversarial input sequence misclassified by the model. We achieved an error rate of $100\%$ on the training set by changing on average $9.18$ words in each of the $2,000$ training reviews. Reviews are on average $71.06$ word long. For instance, we change the review ``I wouldn't rent this one even on dollar rental night.'' into the following misclassified adversarial sequence ``Excellent wouldn't rent this one even on dollar rental night.''. The algorithm is inserting words with highly positive connotations in the input sequence to mislead the RNN model.

%\newpage

\subsection{Recurrent Neural Networks with Sequential Output}

This RNN predicts output sequences from input sequences. Although we use symthetic data, sequence-to-sequence models can for instance be applied to forecast financial market trends. 

\begin{figure}[h] 
	\centering
	\includegraphics[width=0.9\columnwidth]{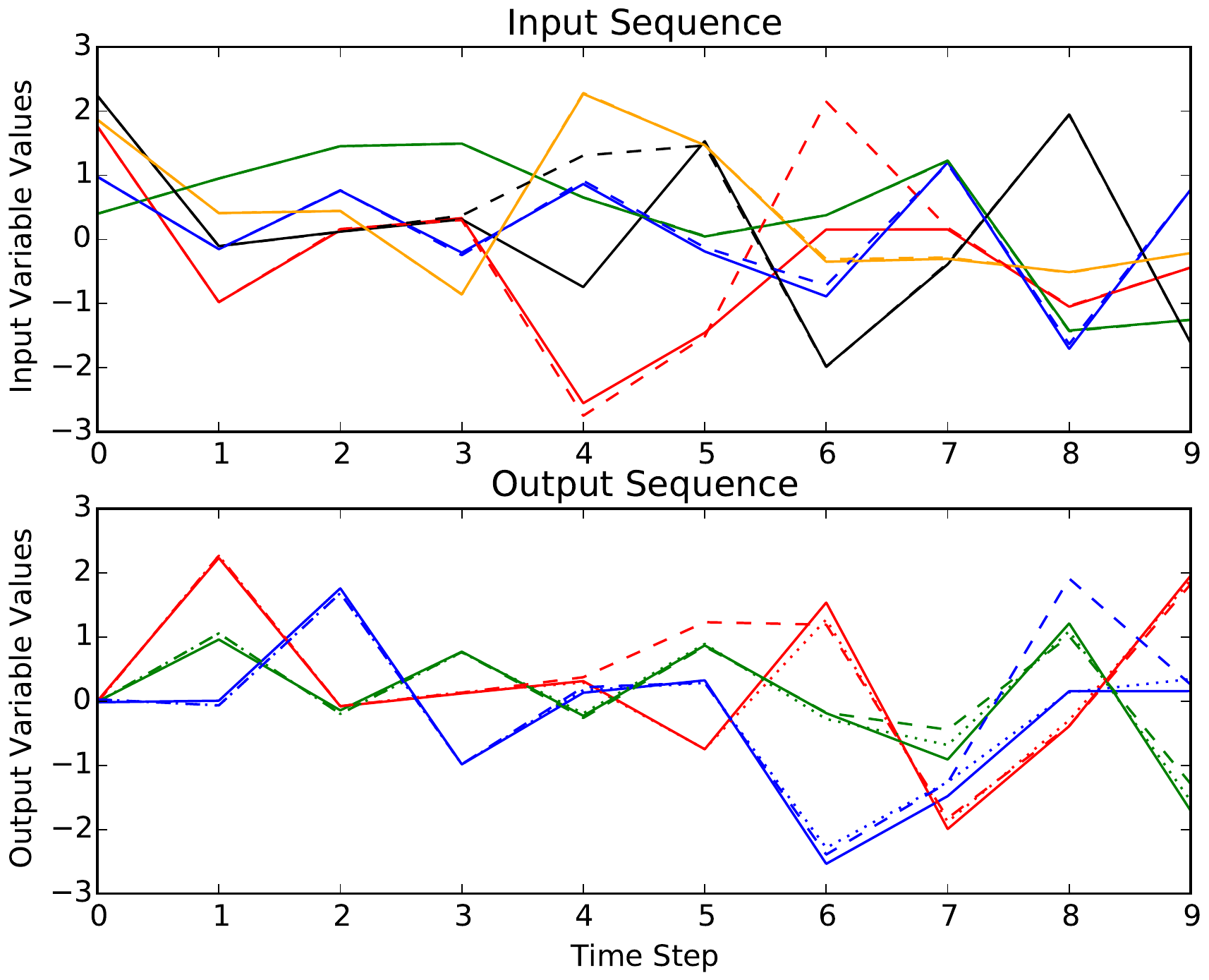} 
	\caption{\textbf{Example input and output sequences of our experimental setup} In the input graph, the solid lines indicate the legitimate input sequence while the dashed lines indicate the crafted adversarial sequence. In the output, solid lines indicate the training target output, dotted lines indicated the model predictions and dashed lines the prediction the model made on the adversarial sequence.}
	\label{fig:experimental-setup-seq-example}
\end{figure}

\vspace*{0.1in}
\noindent\textbf{Experimental Setup -} The sequential RNN is described in Figure~\ref{fig:rnn}. We train on a set of $100$  synthetically generated input and output sequence pairs. Inputs have $5$ values per step and outputs $3$ values per step. Both sequences are $10$ steps long. These values are randomly sampled from a standard normal distribution ($\mu=0$ and  $\sigma^2=1$ for inputs, $\mu=0$ and $\sigma^2=10^{-4}$ for outputs). The random samples are then altered to introduce a strong correlation between a given step of the output sequence and the previous (or last to previous) step of the input sequence. The model is trained for $400$ epochs at a learning rate of $10^{-3}$. The cost is the mean squared error between model predictions and targets. Figure~\ref{fig:experimental-setup-seq-example} shows an example input sequence and the output sequence predicted.

\vspace*{0.1in}
\noindent\textbf{Adversarial Sequences -} We compute the model's Jacobian matrix---which quantifies contributions of each input sequence step to each output sequence step---to craft adversarial sequences. For instance, if we are interested in altering a subset of output steps $\{j\}$, we simply alter the subset of input steps $\{i\}$ with high Jacobian values $J_f[i,j]$ and low Jacobian values $J_f[i,k]$ for $k\neq j$. Figure~\ref{fig:experimental-setup-seq-example} shows example inputs and outputs. Solid lines correspond to the legitimate input sequence and its target output sequence, while (small) dotted lines in the output show model predictions (which closely matches the target). The adversarial sequence---dashed---was crafted to modify value $0$ (red) of step $5$ and value $2$ (blue) of step $8$. It does so by only making important changes in the input sequence at value $3$ (black) of step $4$ and value $0$ (red) of step $6$. Due to space constraints, completing these qualitative results with a detailed quantitative evaluation is left as future work.

\newpage
\section{Discussion and Related Work}
\label{sec:discussion}

This work is part of an active line of research---\emph{adversarial learning}--which studies the behavior of machine learning models trained or deployed in adversarial settings~\cite{barreno2006can}. 

The theoretical approach described in Section~\ref{sec:crafting-adv-sequences} is applicable to any neural network model with recurrent components, independent of its output data type. Our experiments were performed on a LSTM architecture with categorical outputs and a low-dimensional vanilla RNN model with sequential outputs as a preliminary validation of the approach, albeit necessitating additional validation with other RNN model variants, as well as datasets. Future work should also address the grammar of adversarial sequences to improve their semantic meaning and make sure that they are indistinguishable to humans. 

In this paper, we considered a threat model describing adversaries with the capability of accessing the model's architecture---its computational graph---including the values of parameters learned during training. In realistic environments, it is not always possible for adversaries without some type of access to the system hosting the machine learning model to acquire knowledge of these parameters. This limitation has been addressed in the context of deep neural network classifiers by~\cite{papernot2016practical}. The authors introduced a black-box attack for adversaries targeting classifier oracles: the targeted model can be queried for labels with inputs of the adversary's choice. They used a substitute model to approximate the decision boundaries of the unknown targeted model and then crafted adversarial samples using this substitute. These samples are also frequently misclassified by the targeted model due to a property known as \emph{adversarial sample transferability}: samples crafted to be misclassified by a given model are often also misclassified by different models. However, adapting such a black-box attack method to RNNs requires additional research efforts, and is left as future work.

\section{Conclusions}
\label{sec:conclusions}

Models learned using RNNs are not immune from vulnerabilities exploited by adversary carefully selecting perturbations to model inputs, which were uncovered in the context of feed-forward---acyclical---neural networks used for computer vision classification~\cite{szegedy2013intriguing,goodfellow2014explaining,papernot2015limitations}. In this paper, we formalized the problem of crafting adversarial sequences manipulating the output of RNN models. We demonstrated how techniques previously introduced to craft adversarial samples misclassified by neural network classifiers can be adapted to produce sequential adversarial inputs, notably by using computational graph unfolding. In an experiment, we validated our approach by crafting adversarial samples evading models making classification predictions and sequence-to-sequence predictions. 

Future work should investigate adversarial sequences of different data types. As shown by our experiments, switching from computer vision to natural language processing applications introduced difficulties. Unlike previous work, we had to consider the pre-processing of data in our attack. Performing attacks under weaker threat models will also contribute to the better understanding of vulnerabilities and lead to defenses.

\newpage

{\scriptsize

% use section* for acknowledgment
\section*{Acknowledgments}

\noindent Research was sponsored by the Army Research Laboratory (ARL) and was accomplished
under Cooperative Agreement Number W911NF-13-2-0045 (ARL Cyber Security
CRA). The views and conclusions contained in this document are those of the
authors and should not be interpreted as representing the official policies,
either expressed or implied, of the ARL or the U.S.
Government. The U.S. Government is authorized to reproduce and distribute
reprints for Government purposes notwithstanding any copyright notation.

}

% Generated by IEEEtran.bst, version: 1.13 (2008/09/30)

% that's all folks
\end{document}